\documentclass{iau} 
\usepackage{graphicx}

\title[Globular Cluster Streams] 
{Globular Cluster Streams as Galactic High-Precision Scales}

\author[K\"upper et al.]   
{Andreas H.W. K\"upper$^{1}$\thanks{Hubble Fellow}, Eduardo Balbinot$^2$, Ana Bonaca$^3$, Kathryn V. Johnston$^1$, David W. Hogg$^4$, Pavel Kroupa$^5$, and Basilio X. Santiago$^{6,7}$
}

\affiliation{
$^1$Department of Astronomy, Columbia University, \\550 West 120th Street, New York, NY 10027, USA \\email: {\tt akuepper@astro.columbia.edu}
\\[\affilskip]
$^2$Department of Physics, University of Surrey, \\Guildford GU2 7XH, UK\\[\affilskip]
$^3$Department of Astronomy, Yale University, \\New Haven, CT 06511, USA\\[\affilskip]
$^4$Center for Cosmology and Particle Physics, Department of Physics, New York University, \\4 Washington Place,
New York, NY 10003, USA\\[\affilskip]
$^5$Helmholtz-Institut f\"ur Strahlen- und Kernphysik (HISKP), University of Bonn, \\Nussallee 14-16, 53115 Bonn, Germany\\[\affilskip]
$^6$Departamento de Astronomia, Universidade Federal do Rio Grande do Sul, \\Av. Bento Gon\c{c}alves 9500, Porto Alegre 91501-970, RS, Brasil\\[\affilskip]
$^7$Laborat\'orio Interinstitucional de e-Astronomia - LIneA, \\Rua Gal. Jos\'e Cristino 77, Rio de Janeiro, RJ - 20921-400, Brasil
}

\pubyear{2015}
\volume{317}  
\jname{The General Assembly of Galaxy Halos: Structure, Origin and Evolution}
\editors{A. Bragaglia, M. Arnaboldi, M. Rejkuba, D. Romano, eds.}
\begin{document}

\maketitle

\begin{abstract}
Tidal streams of globular clusters are ideal tracers of the Galactic gravitational potential. Compared to the few known, complex and diffuse dwarf-galaxy streams, they are kinematically cold, have thin morphologies and are abundant in the halo of the Milky Way. Their coldness and thinness in combination with potential epicyclic substructure in the vicinity of the stream progenitor turns them into high-precision scales. With the example of Palomar\,5, we demonstrate how modeling of a globular cluster stream allows us to simultaneously measure the properties of the disrupting globular cluster, its orbital motion, and the gravitational potential of the Milky Way.

\end{abstract}

\firstsection 
\section{Why thin globular cluster streams are so valuable}
Within the past decade, the number of wide-field imaging surveys and spectroscopic campaigns has grown exponentially. In the vast amount of deep, high-quality data that has become available, a multitude of thin and cold stellar streams has been discovered. Due to their faintness, all of these thin streams were found in the halo of the Milky Way, and most probably originate from disrupting or disrupted globular clusters (e.g., \cite[Bonaca, Geha \& Kallivayalil 2012]{Bonaca12}, \cite[Grillmair et al.~2013]{Grillmair13}, \cite[Bernard et al.~2014]{Bernard14}, \cite[Koposov et al.~2014]{Koposov14}).

Similar to the longer, but more diffuse, dwarf galaxy streams like the Sagittarius stream (e.g., \cite[Johnston et al.~2005]{Johnston05}, \cite[Law \& Majewski 2010]{Law10}), thin globular cluster streams (GCS) are valuable tracers of the Galactic gravitational potential (\cite[Bonaca et al.~2014]{Bonaca14}). Their coherence in phase space makes them ideal instruments for measuring the mass and shape of the otherwise invisible dark halo of the Galaxy, as has been demonstrated by \cite[Koposov et al.~(2009)]{Koposov09} with the example of GD-1.

But the pure existence of thin streams in the halo of the Milky Way tells us even more: \cite[Pearson et al.~(2015)]{Pearson15} found that GCS are very sensitive to the non-sphericity of the gravitational potential of their host galaxy. The authors showed that triaxial halo configurations can cause stream fanning -- a broadening and diffusion of the stream perpendicular to the orbital motion of the satellite. Stream fanning makes the already faint and cold GCS harder to detect in imaging surveys, as their surface density is pushed beyond the detection limit. In a systematic study of orbits within a triaxial galaxy potential, \cite[Price-Whelan et al.~(2015)]{Price15} demonstrated that thin streams can only occupy specific (regular) regions of orbital space. The existence of the many observed thin streams will, therefore, tell us something about the shape of the Milky Way's gravitational potential.

Furthermore, kinematically cold GCS are powerful potential probes as they inhibit dynamical substructure caused by apparent epicyclic motion of stars escaping the gravitational potential of the cluster (\cite[K\"upper, MacLeod \& Heggie 2008]{Kupper08}, \cite[Just et al.~2009]{Just09}). This substructure can be well understood, as it solely depends on the mass of the globular cluster, its orbital motion, and the shape of the galactic gravitational potential (\cite[K\"upper, Lane \& Heggie 2012]{Kupper12}). Long, thin streams of globular clusters that exhibit substructure are therefore high-precision scales of the host galaxy potential. The high achievable precision is due to the unique properties of GCS that let us accurately constrain their progenitor's stellar mass and orbit within the Galaxy:
\begin{enumerate}
\item since globular clusters have simple compositions compared to dwarf galaxies, the progenitors of GCS can be well characterized. Mass estimates are accurate up to the uncertainties of globular cluster mass-to-light ratios, i.e., about a factor of two,
\item the simple stellar compositions of globular clusters, furthermore, allow for a clearer separation of GCS stars from fore- and background contaminations,   
\item the relatively small widths of GCS enables a precise location of the streams, and
\item their cold compositions (i.e., velocity spread among stream members) allows for accurate velocity information along the GCS.
\end{enumerate}

GCS with epicyclic overdensities therefore contain a lot of different information (cluster mass, orbital motion, host gravitational potential), which has to be decoded and disentangled via modeling. The more information is available on the mass of the respective globular cluster, its orbital motion or the host gravitational potential, the better we can constrain the other components. We demonstrated this for the Milky Way globular cluster Palomar\,5.

\section{Modeling Palomar\,5 and its tidal stream}

\begin{figure}[b]
\begin{center}
 \includegraphics[width=5.4in]{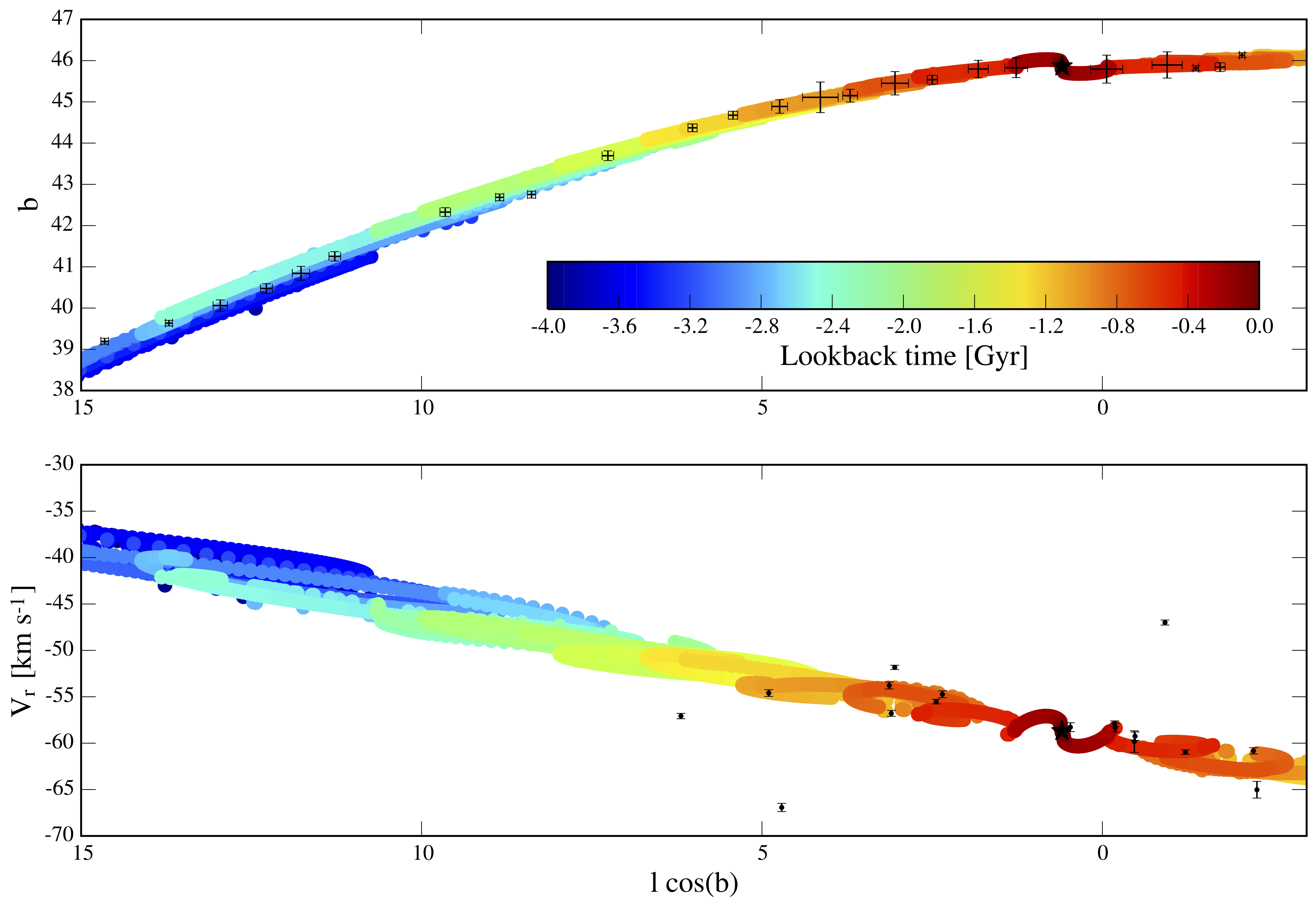} 
 \caption{Streakline model of Pal\,5 consisting of $2\times4000$ test particles that were released from the progenitor in intervals of 1\,Myr. The color coding in both panels shows their release times. Upper panel: projection of the stream on the sky. Black data points are locally over-dense surface density regions of Pal\,5-like stars in SDSS data. Lower panel: radial velocity gradient along the stream with measured velocities of red giants lying in projection within the stream from Odenkirchen et al.~(2009).}
   \label{fig1}
\end{center}
\end{figure}

The Milky Way globular cluster Palomar\,5 (Pal\,5) shows a thin, $>20$\,deg long tidal stream, which was first discovered by  \cite[Odenkirchen et al.~(2001)]{Odenkirchen01} in commissioning data of the Sloan Digital Sky Survey (SDSS). A detailed review of the available observational data on Pal\,5 and its stream can be found in \cite[K\"upper et al.~(2015)]{Kupper15}. 

In this publication, we extensively modeled Pal\,5 and its tidal stream. We used a difference-of-Gaussian procedure to detect the densest and most significant regions within the Pal\,5 stream. This ansatz also allowed us to locate potential epicyclic overdensities within the prominent tidal stream. We combined this surface density information with radial velocity measurements along the stream from \cite[Odenkirchen et al.~(2009)]{Odenkirchen09}. Both over-dense regions and radial velocity measurements are shown in Fig.~\ref{fig1}.

Similar to the \textsc{Fast Forward} method developed in \cite[Bonaca et al.~(2014)]{Bonaca14}, we used streakline models of the Pal\,5 stream to evaluate the likelihood of a given set of model parameters. Our stream models encompassed 10 free parameters, describing the progenitor globular cluster (mass, mass-loss rate, distance, proper motion), the position and motion of the Sun, and the properties of the Galactic gravitational potential (mass, size, and $q_z$ -- the flattening perpendicular to the Galactic disk). Posterior probability distributions of the parameters were obtained through Markov-chain Monte Carlo sampling using the freely available code \textit{emcee} (\cite[Foreman et al.~2013]{Foreman13}).

We were able to tightly constrain all 10 model parameters and give uncertainties for each one, which demonstrates the power of substructured GCS when modeled with streaklines in a Bayesian framework. For example, we found the shape of the Galactic halo potential within the inner 19\,kpc with a value of $q_z = 0.95^{+0.16}_{-0.12}$ close to being spherical. \cite[Koposov et al.~(2009)]{Koposov09} came to a similar conclusion fitting orbits to the significantly longer, but fainter GD-1 stream. With three free model parameters, they were able to rule out a halo flattening smaller than $q_z = 0.89$ with 90\% confidence, but could not give an upper limit on its possible prolateness. Similarly, results from modeling of the long and diffuse Sagittarius dwarf galaxy stream vary between strongly prolate \cite[(Helmi 2004)]{Helmi04} and oblate \cite[(Johnston et al.~2005)]{Johnston05}, both without uncertainty estimates. Our modeling of Pal\,5 is therefore a significant improvement over previous investigations.

\section{Outlook}

\begin{figure}[b]
\begin{center}
 \includegraphics[width=5.4in]{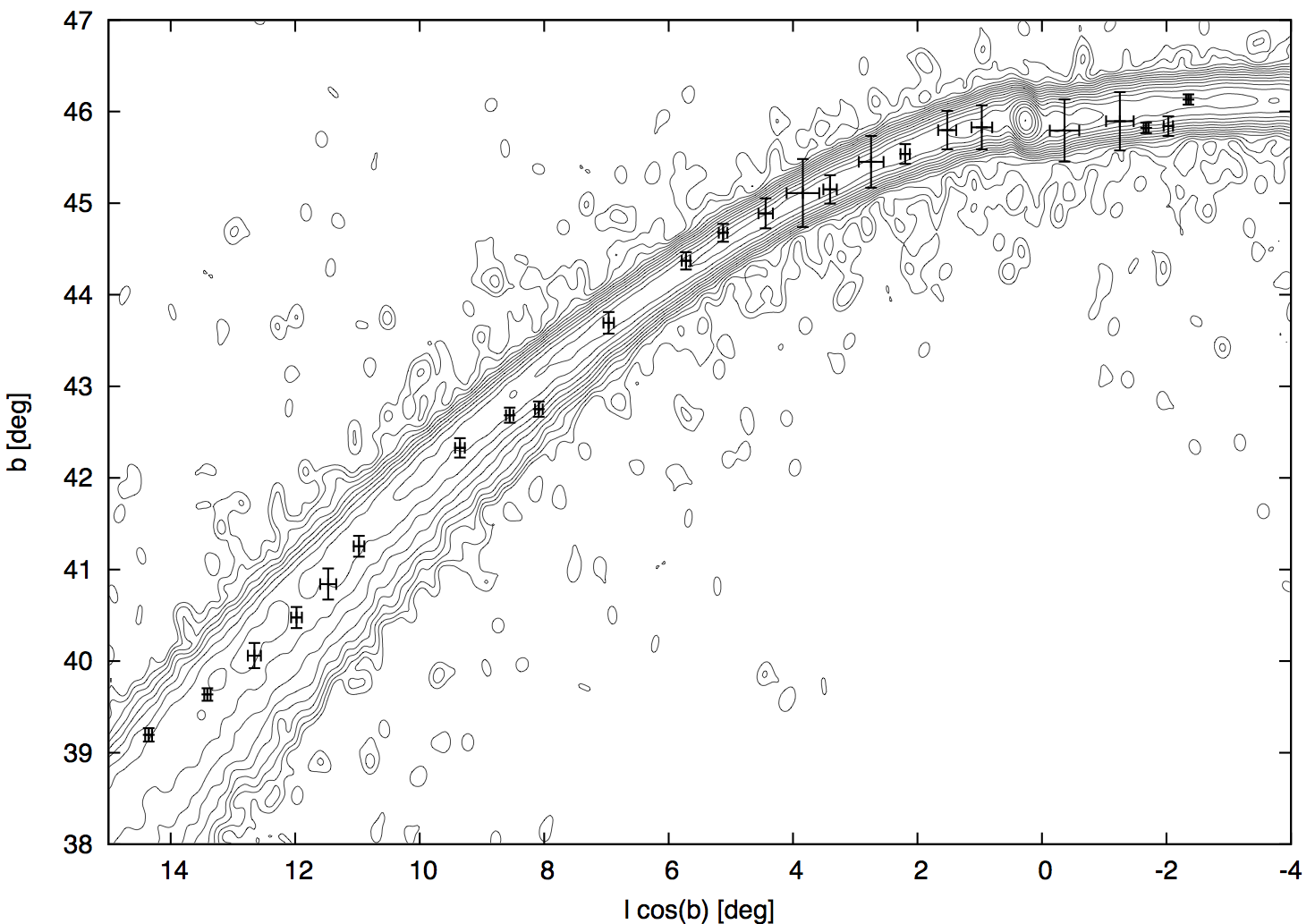} 
 \caption{Contour density map of one of our first $N$-body models of Pal\,5. Contour levels are equally spaced in log density. Black data points are the same as in Fig.~\ref{fig1}. Like the streakline model, the $N$-body stream is thin along the whole extent, and is most collimated at about $l\cos(b)=4$\,deg, where the stream shows its most pronounced overdensities in SDSS data.}
   \label{fig2}
\end{center}
\end{figure}

New observational data on Pal\,5 and its stream will help to inform and improve further modeling of the stream. For example, \cite[Kuzma et al.~(2014)]{Kuzma14} published 39 additional radial velocity measurements of red giants along the Pal\,5 stream. Moreover, \cite[Fritz \& Kallivayalil (2015)]{Kallivayalil15} obtained 15 year baseline proper motions of Pal\,5 from SDSS data in combination with LBT imaging data. Further data from, e.g., HST, Spitzer, Keck and CTIO will give us an even deeper and more detailed look on the Pal\,5 stream and its kinematics in the near future.

But we can already use the constraints from our current streakline modeling on Pal\,5's orbit within the Galactic tidal field to verify our understanding of the destruction of the cluster. Using the parameters of our best-fit models, we are currently following up our simplified stream modeling with accurate $N$-body models of the disrupting Pal\,5. We aim at extending previous investigations on Pal\,5's evolution by \cite[Dehnen et al.~(2004)]{Dehnen04} and \cite[Mastrobuono-Battisti et al.~(2012)]{Mastrobuono12}, to link the internal evolution of the cluster to the appearance of its tidal stream.

Our first $N$-body models with the GPU-enabled, direct $N$-body code \textsc{Nbody6} (\cite[Aarseth 2003]{Aarseth03}) show good agreement with the overall properties of the streakline models (Fig.~\ref{fig2}), but also give us new exciting challenges. Internal properties of Pal\,5, such as its half-light radius, are observationally well determined and will give us further constraints on the evolution of Pal\,5 (cf., \cite[Zonoozi et al.~2011]{Zonoozi11}; \cite[Zonoozi et al.~2014]{Zonoozi14}). Ultimately, full $N$-body modeling of Pal\,5 and its stream will give us the unique opportunity to understand the birth, life and death of a globular cluster in unprecedented detail.


\begin{thebibliography}{}

\bibitem[Aarseth (2003)]{Aarseth03}
{Aarseth, S.~J.} 2003, \textit{Gravitational N-Body Simulations} (Cambridge University Press)

\bibitem[Bernard \etal\ (2014)]{Bernard14}
{Bernard, E.~J., Ferguson, A.~M.~N., Schlafly, E.~F., \etal\ } 2014, \textit{MNRAS}, 443, L84

\bibitem[Bonaca \etal\ (2012)]{Bonaca12}
{Bonaca, A., Geha, M., Kallivayalil, N.} 2012, \textit{ApJ} (Letters), 760, L6

\bibitem[Bonaca \etal\ (2014)]{Bonaca14}
{Bonaca, A., Geha, M., K{\"u}pper, A.~H.~W., Diemand, J., Johnston, K.~V., Hogg, D.~W.} 2014, \textit{ApJ}, 795, 94

\bibitem[Dehnen \etal\ (2004)]{Dehnen04}
{Dehnen, W., Odenkirchen, M., Grebel, E.~K., Rix, H.-W.} 2004, \textit{AJ},  127, 2753

\bibitem[Foreman-Mackey \etal\ (2013)]{Foreman13}
{Foreman-Mackey, D., Hogg, D.~W., Lang, D., Goodman, J.} 2013, \textit{PASP},  125, 306

\bibitem[Fritz \& Kallivayalil (2015)]{Fritz15}
{Fritz, T.~K., Kallivayalil, N.} 2015, accepted for publication in \textit{ApJ}, arXiv:1508.06647

\bibitem[Grillmair \etal\ (2013)]{Grillmair13}
{Grillmair, C.~J., Cutri, R., Masci, F.~J., Conrow, T., Sesar, B., Eisenhardt, P.~R.~M., Wright, E.~L.} 2013, \textit{ApJ} (Letters), 769, L23

\bibitem[Helmi (2004)]{Helmi04} 
{Helmi, A.} 2004, \textit{ApJ}, 610, L97 

\bibitem[Johnston \etal\ (2005)]{Johnston05} 
{Johnston, K.~V., Law, D.~R., Majewski, S.~R.} 2005, \textit{ApJ}, 619, 800 

\bibitem[Just \etal\ (2009)]{Just09}
{Just, A., Berczik, P., Petrov, M.~I., Ernst, A.} 2009, \textit{MNRAS}, 392, 969

\bibitem[Koposov \etal\ (2010)]{Koposov10}
{Koposov, S.~E., Rix, H.-W., Hogg, D.~W.} 2010, \textit{ApJ}, 712, 260

\bibitem[Koposov \etal\ (2014)]{Koposov14}
{Koposov, S.~E., Irwin, M., Belokurov, V., \etal\ } 2014, \textit{MNRAS}, 442, L85

\bibitem[K{\"u}pper \etal\ (2008)]{Kupper08}
{K\"upper, A.~H.~W., MacLeod, A., Heggie, D.~C.} 2008, \textit{MNRAS}, 387, 1248

\bibitem[K\"upper \etal\ (2012)]{Kupper12}
{K\"upper, A.~H.~W., Lane, R.~R., Heggie, D.~C.} 2012, \textit{MNRAS}, 420, 2700

\bibitem[K\"upper \etal\ (2015)]{Kupper15}
{K{\"u}pper, A.~H.~W., Balbinot, E., Bonaca, A., Johnston, K.~V., Hogg, D.~W., Kroupa, P., Santiago, B.~X.} 2015,
\textit{ApJ}, 803, 80 

\bibitem[Kuzma \etal\ (2015)]{Kuzma14}
{Kuzma, P.~B., Da Costa, G.~S., Keller, S.~C., Maunder, E.} 2015, \textit{MNRAS}, 446, 3297

\bibitem[Law \& Majewski (2010)]{Law10} 
{Law, D.~R., Majewski, S.~R.} 2010, \textit{ApJ}, 714, 229 

\bibitem[Mastrobuono-Battisti \etal\ (2012)]{Mastrobuono12}
{Mastrobuono-Battisti, A., Di Matteo, P., Montuori, M., Haywood, M.} 2012, \textit{A\&A}, 546, L7

\bibitem[Odenkirchen \etal\ (2001)]{Odenkirchen01} 
{Odenkirchen, M., Grebel, E.~K., Rockosi, C.~M., \etal\ } 2001, \textit{ApJ} (Letters), 548, L165

\bibitem[Odenkirchen \etal\ (2009)]{Odenkirchen09}
{Odenkirchen, M., Grebel, E.~K., Kayser, A., Rix, H.-W., Dehnen, W.} 2009, \textit{AJ}, 137, 3378

\bibitem[Pearson \etal\ (2015)]{Pearson15}
{Pearson, S., K{\"u}pper, A.~H.~W., Johnston, K.~V., Price-Whelan, A.~M.} 2015, \textit{ApJ}, 799, 28

\bibitem[Price-Whelan et al.~(2015)]{Price15} 
{Price-Whelan, A.~M., Johnston, K.~V., Valluri, M., Pearson, S., K\"upper, A.~H.~W., Hogg, D.~W.} 2015, submitted to \textit{ApJ}, arXiv:1507.08662 

\bibitem[Zonoozi \etal\ (2011)]{Zonoozi11} 
{Zonoozi, A.~H., K{\"u}pper, A.~H.~W., Baumgardt, H., Haghi, H., Kroupa, P., Hilker, M.} 2011, \textit{MNRAS}, 411, 1989 

\bibitem[Zonoozi \etal\ (2014)]{Zonoozi14} 
{Zonoozi, A.~H., Haghi, H., K{\"u}pper, A.~H.~W., Baumgardt, H., Frank, M.~J., Kroupa, P.} 2014, \textit{MNRAS}, 440, 3172 

\end{thebibliography}
\end{document}